\documentclass[fleqn,twoside]{article}
\usepackage{espcrc2}

\usepackage{graphicx}
\usepackage[figuresright]{rotating}

\newcommand{\AmS}{{\protect\the\textfont2
  A\kern-.1667em\lower.5ex\hbox{M}\kern-.125emS}}

\title{Hadronic Spectral Functions above the QCD Phase Transition
\thanks{Presented by M. Asakawa}}

\author{M. Asakawa\address{Department of Physics, 
        Kyoto University, Kyoto 606-8502, Japan},
        T. Hatsuda\address{Department of Physics,
        University of Tokyo, Tokyo 113-0033, Japan} 
        and
        Y. Nakahara\address{Department of Physics, Nagoya University, 
        Nagoya, 464-8602, Japan}}
       
\begin{document}

\begin{abstract}
We extract the spectral functions in the scalar, pseudo-scalar,
vector, and axial-vector channels above the deconfinement
phase transition temperature ($T_c$) using the maximum entropy method (MEM).
We use anisotropic lattices, $32^3 \times 32$, 40, 54, 72, 80, and 96
(corresponding to $T= 2.3 T_c \rightarrow 0.8 T_c$),
with the renormalized anisotropy $\xi=4.0$ to
have enough temporal data points to carry out the MEM analysis.
Our result suggests that the spectral functions continue to possess
non-trivial structures even above $T_c$ and in addition 
that there is a qualitative change in the state of the deconfined 
matter between $1.5T_c$ and $2T_c$.
\vspace{1pc}
\end{abstract}

\maketitle

\section{INTRODUCTION}

The spectral functions (SPFs) of hadronic operators play
an important role in QCD. The modification of hadrons, which
has been suggested, for example, by the dilepton production
enhancement in high energy nuclear collisions observed
at CERN SPS, can be formulated in terms of SPFs. 
Although there have been numerous theoretical attempts
to understand the modification of SPFs
at finite temperature and/or density \cite{modification},
the exact nature of the hadronic modes in matter is not 
understood well.
 
Recently, the maximum entropy method (MEM) has been used
to extract SPFs from lattice QCD data for the first time by the 
present authors \cite{nah99,ahn01}. 
The Euclidean 2-point functions in the temporal direction
and the associated SPFs are related by the Laplace transform.
On the lattice, only a finite number of data points with statistical
noise are available in the temporal direction. Therefore, 
direct Laplace inversion from lattice data to SPF 
is an ill-posed problem. MEM is a method
to evade such difficulty on the basis of the Bayes' theorem
in the theory of statistical inference \cite{ahn01}. 

\section{HOW MANY DATA POINTS ARE NECESSARY FOR MEM ?}

As we have discussed in detail in \cite{ahn01}, the result of
MEM depends strongly on the number of temporal data points.
The more data points are used, the closer the MEM result is to the true SPF. 
Moreover, there is the minimum number of data points 
$N$ to perform the reliable MEM analysis.

\begin{figure}[thb]
\begin{flushleft}
\includegraphics[width=1.0 \linewidth]{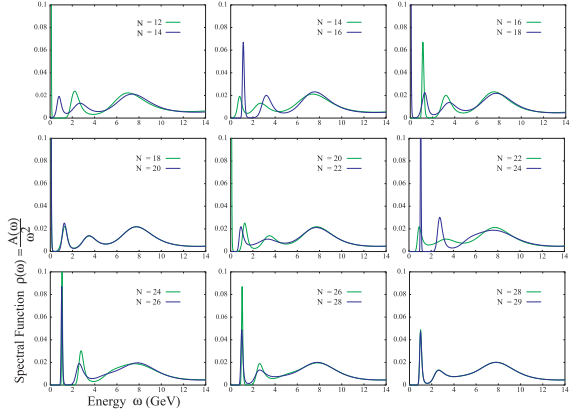}
\end{flushleft}
\vspace{-13mm}
\caption{MEM results in the vector channel 
with 10 different $N$'s. 
 }
\label{ntchange}
\end{figure}

In order to find this minimum $N$ at $T=0$ on the lattice,
we have carried out the following analysis \cite{ahn02}.
First we calculated hadronic correlators 
with the quenched approximation on an isotropic
$40^3 \times 30$ lattice at $\beta = 6.47$ \cite{ahn02,cppacs}.
The number of gauge configurations is 160. 
Then, MEM analysis has been done for
the vector channel by using $N$ data points 
(at $\tau = 2, \cdots, N / 2 +1$ and $31 - N/2, \cdots, 30$)
out of the total 30 data points. The result is shown in Fig. \ref{ntchange}.
The figure clearly shows that, SPF changes considerably
as $N$ and that at least about 30 data points are necessary
for the convergence of the result. This minimum $N$ 
would depend on $\beta$, and on whether one employs
anisotropic lattices, improved actions, and so on.
Nevertheless, we use this number as a practical guide in the following.

\section{ANISOTROPIC LATTICE AT $T\neq 0$}

At finite temperature ($T$), 
because of the relation $T=1/N_{\tau} a_{\tau}$ 
($N_{\tau}$ and $a_\tau$ being the 
temporal lattice size and spacing, respectively),
less and less data points are available as $T$ increases
if $a_\tau$ is fixed.
Requiring that $N_{\tau} > 30$ holds
even at the highest $T$, e.g. $2.5T_c$,
we are inevitably led to use an anisotropic lattice. 

We have used the bare anisotropy $\xi_0 = 3.5$ and $\beta=7.0$
with the naive plaquette action. For the quark part, 
the standard Wilson action with the quenched approximation is used.
The corresponding renormalized anisotropy is
$\xi=4.0$ \cite{karsch_aniso} and the lattice spacing
is $a_\tau = 1/4 \cdot a_{\sigma} = 9.75\times 10^{-3}$ fm.
Simulations are done on $32^3\times N_\tau $ lattices 
with $N_\tau = 32,$ 40, 54, 72, 80, and 96. This corresponds to 
$T \simeq 2.3T_c$, $1.9T_c$, $1.4T_c$, $1.04T_c$, $0.93T_c$, and $0.78T_c$,
respectively. More than 100 configurations are generated
for each $N_\tau$. The details of the lattice parameters
used in the calculation will be given in \cite{ahn02}.

We have studied 2-point correlation functions in the 
scalar (S), pseudo-scalar (PS), vector (V), and axial-vector (AV)
channels. The default models, motivated by perturbative QCD, in each
channel are $m=0.60$, 1.15, 0.40, and 0.35, respectively.
\begin{figure}[thb]
\begin{flushleft}
\includegraphics[width=0.9 \linewidth]{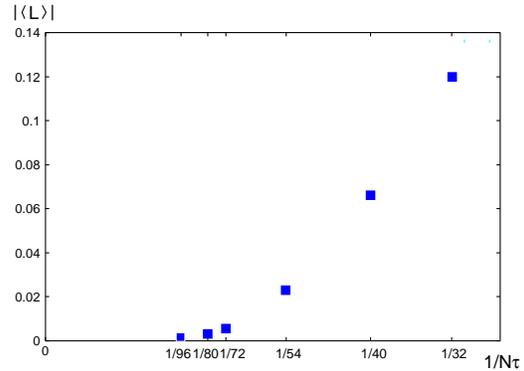}
\end{flushleft}
\vspace{-13mm}
\caption{Polyakov loop expectation value as a function
of $N_\tau$.}
\label{polyakov}
\end{figure}
The lattice data points used in our MEM analysis are
$\tau = 5, \cdots, 21$ and $N_\tau-19, \cdots, N_\tau-3$
for $N_\tau = 54$, 72, 80, and 96. For $N_\tau=32$ and 40,
$\tau = 5, \cdots, N_\tau-3$ are used. When the source and
sink are closer than $\xi a_{\tau}$ in the temporal direction, 
artifact could arise due to unphysical modes at
$\omega \ge \pi/\xi a_\tau$. Therefore, we leave out
the data points near the edges, i.e., $\tau = 1, \cdots, 4$ and
$N_\tau- 2, \cdots, N_\tau$. 
 
The number of lattice data points used in the MEM analysis
is fixed almost the same at each $N_\tau$ except at
$N_\tau = 32$ to keep the resolution in the MEM analysis
unchanged. The rest of the MEM procedures such as the averaging over $\alpha$
are basically the same as described in \cite{ahn01} and will be
presented in detail in \cite{ahn02}.

Fig. \ref{polyakov} shows the Polyakov loop expectation value
$|\langle L \rangle |$ at each $N_\tau$. It starts to deviate from
zero around $N_\tau = 80 (72)$, which corresponds to $T_c = 253 (281)$ MeV,
and is comparable to the real value 271 $\pm$ 2 MeV.
We have checked that the Polyakov loop susceptibility
also shows a peak in this region. In the following,
we use the number $T_c = 271$ MeV.

\section{RESULTS OF MEM}
 
\begin{figure}[thb]
\begin{flushleft}
\includegraphics[width=0.9 \linewidth]
{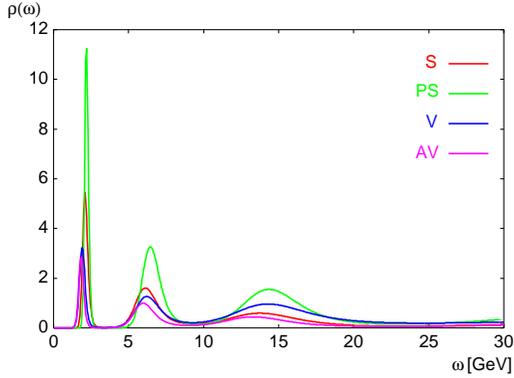}
\end{flushleft}
\vspace{-13mm}
\caption{Spectral Functions for $N_\tau = 54$
 ($T\simeq 1.4T_c$)}
\label{SPF_Nt54}
\end{figure}

In Fig. \ref{SPF_Nt54},
we show the results of the MEM analysis for SPF
$\rho(\omega)$ in the S, PS, V, and AV channels on
$32^3 \times 54$ lattice 
($T\simeq 1.4T_c$).
In this calculation, we have used the hopping parameter
corresponding to $m_\pi/m_\rho \simeq 0.7$ at $T=0$.

If the quark-gluon plasma is such an object as described as a free gas
of massive quarks, anti-quarks and gluons, $\rho(\omega)$ would show
a flat behavior with a smooth rise from zero above the $q\bar{q}$ threshold 
as a function of $\omega$. To the contrary, Fig. \ref{SPF_Nt54}
shows that SPFs possess nontrivial structures
even above $T_c$ in all four channels. There are a sharp peak
at about $\omega = 2$ GeV and two bumps above the peak.
Also SPFs in all channels almost degenerate, which shows a strong 
signature of the restoration of chiral symmetry.
The peak and bump structures, 
and the suppression of SPFs below $\omega = 2$ GeV 
are statistically significant according to our error analysis.
We have used $N_\tau$ twice as large as that
employed in \cite{bielefeld_spf}. As shown in 
Sec.2, this is necessary for reliable MEM analysis.

\begin{figure}[thb]
\begin{flushleft}
\includegraphics[width=0.9 \linewidth]{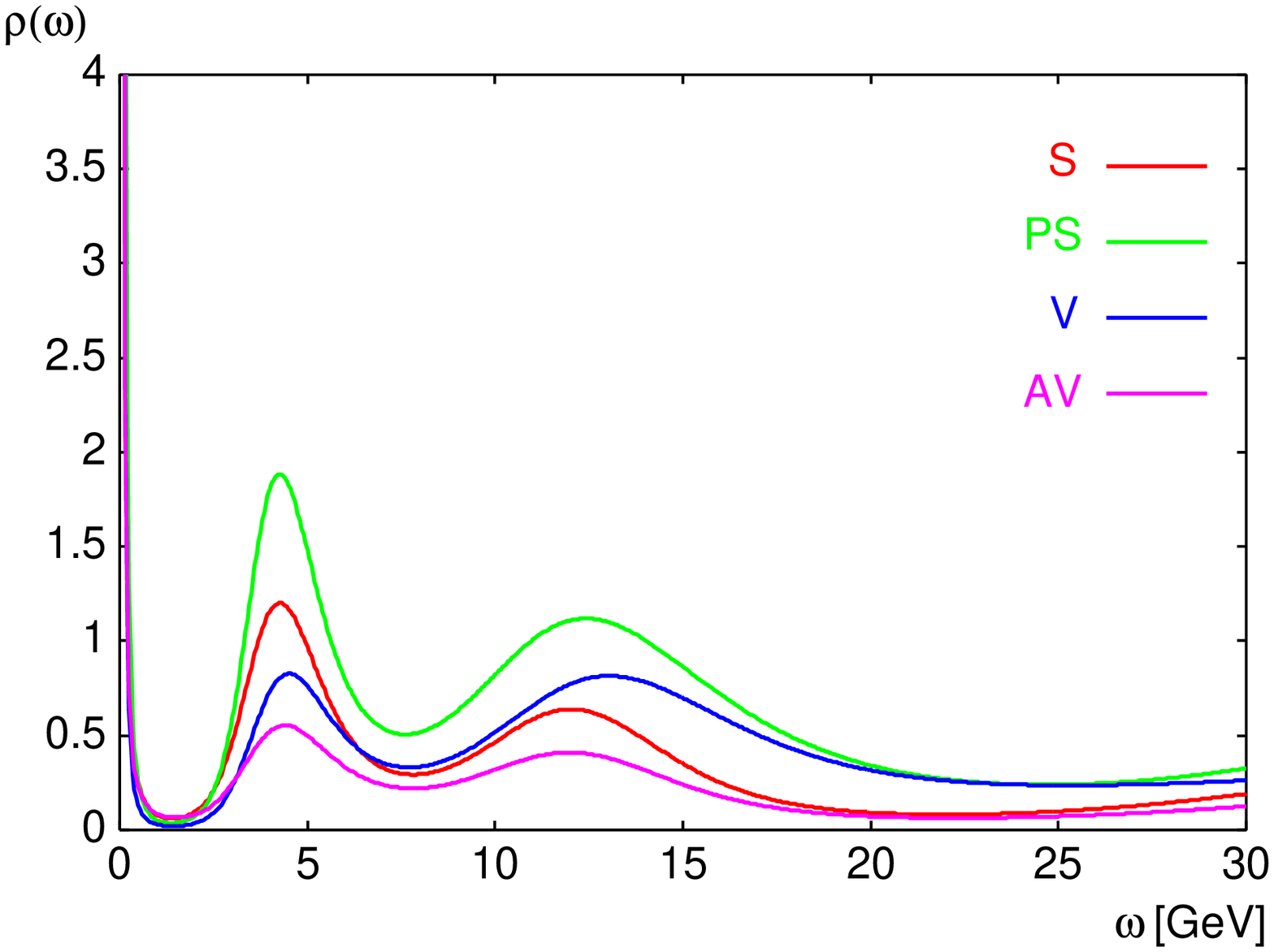}
\end{flushleft}
\vspace{-13mm}
\caption{Spectral Functions for $N_\tau = 40$
 ($T\simeq 1.9 T_c$)} 
\label{SPF_Nt40}
\end{figure}

In Fig. \ref{SPF_Nt40},
we show the results at $N_\tau = 40$,
i.e., $T\simeq 1.9T_c$.
There is an apparent peak around $\omega = 0$.
A possible explanation of the peak
is the effect of Landau damping \cite{HKL93}.
However, this peak is, at the moment, not statistically significant yet.
The peaks around $\omega \simeq 4.5$ GeV are significantly
broader than those around 2 GeV 
at $T\simeq 1.9 T_c$. 

We have carried out an
error analysis for the average of the
SPFs in small regions around the peaks and
found that the broadening is indeed statistically significant.
SPFs in the case of $N_\tau = 32$ 
$(T\simeq 2.3T_c)$ have similar structures as those at
$T\simeq 1.9 T_c$ except the shift
of the peak positions. This suggests a possibility of a qualitative 
change in the state of the deconfined phase between $1.5T_c$ and $2T_c$. 

{\bf Acknowledgement} This work was supported in part by
the Grants-in-Aid by Ministry of Education 
(No. 12640263, No. 12640296, and No. 14540255).
Lattice calculations have been carried out with the
CP-PACS computer under the ``Large-scale Numerical Simulation Program''
of Center for Computational Physics, University of Tsukuba.

\end{document}